\documentclass[twocolumn,floatfix,superscriptaddress,showpacs,showkeys]{revtex4}

\usepackage{amssymb}
\usepackage{amsmath}
\usepackage{amsfonts}
\usepackage{epsfig}
\usepackage{subfigure}
\usepackage{graphics}
\usepackage{float}
\usepackage{latexsym}

\begin{document}
\title{Survey of $A_{LT'}$ asymmetries in semi-exclusive electron scattering \\
on $^{4}$He and $^{12}$C}

%%%%%%%%%%%% New author list %%%%%%%%%%%%%%%%%%%%%%%%

%%%%%%%%%%%%%%%%%%%%%%%%%%%%%%%%%%%%%%%%%%%% 
%  Photon only people have been excluded 
%  Electron only data assumed 
%%%%%%%%%%%%%%%%%%%%%%%%%%%%%%%%%%%%%%%%%%%% 
%%%%%%%%%%%% Institutes Number and defintions %%%%%%%%%%%
 
% FIRST time through establishes the order
%%%%%%%%%%%%%%%%%%%%%%%%%%%%%%%%%%% 
\newcommand*{\ASU }{ Arizona State University, Tempe, Arizona 85287-1504} 
\affiliation{\ASU } 

\newcommand*{\SACLAY }{ CEA-Saclay, Service de Physique Nucl\'eaire, F91191 Gif-sur-Yvette,Cedex, France} 
\affiliation{\SACLAY } 

\newcommand*{\UCLA }{ University of California at Los Angeles, Los Angeles, California  90095-1547} 
\affiliation{\UCLA } 

\newcommand*{\CMU }{ Carnegie Mellon University, Pittsburgh, Pennsylvania 15213} 
\affiliation{\CMU } 

\newcommand*{\CUA }{ Catholic University of America, Washington, D.C. 20064} 
\affiliation{\CUA } 

\newcommand*{\CNU }{ Christopher Newport University, Newport News, Virginia 23606} 
\affiliation{\CNU } 

\newcommand*{\UCONN }{ University of Connecticut, Storrs, Connecticut 06269} 
\affiliation{\UCONN } 

\newcommand*{\DUKE }{ Duke University, Durham, North Carolina 27708-0305} 
\affiliation{\DUKE } 

\newcommand*{\ECOSSEE }{ Edinburgh University, Edinburgh EH9 3JZ, United Kingdom} 
\affiliation{\ECOSSEE } 

\newcommand*{\FIU }{ Florida International University, Miami, Florida 33199} 
\affiliation{\FIU } 

\newcommand*{\FSU }{ Florida State University, Tallahassee, Florida 32306} 
\affiliation{\FSU } 

\newcommand*{\GENT}{Department of Subatomic- and Radiation Physics, 
Ghent University, Belgium}
\affiliation{\GENT}

\newcommand*{\GEISSEN }{ Physikalisches Institut der Universitaet Giessen, 35392 Giessen, Germany} 
\affiliation{\GEISSEN } 

\newcommand*{\GWU }{ The George Washington University, Washington, DC 20052} 
\affiliation{\GWU } 

\newcommand*{\ECOSSEG }{ University of Glasgow, Glasgow G12 8QQ, United Kingdom} 
\affiliation{\ECOSSEG } 

\newcommand*{\INFNFR }{ INFN, Laboratori Nazionali di Frascati, Frascati, Italy} 
\affiliation{\INFNFR } 

\newcommand*{\INFNGE }{ INFN, Sezione di Genova, 16146 Genova, Italy} 
\affiliation{\INFNGE } 

\newcommand*{\ORSAY }{ Institut de Physique Nucleaire ORSAY, Orsay, France} 
\affiliation{\ORSAY } 

\newcommand*{\ITEP }{ Institute of Theoretical and Experimental Physics, Moscow, 117259, Russia} 
\affiliation{\ITEP } 

\newcommand*{\JMU }{ James Madison University, Harrisonburg, Virginia 22807} 
\affiliation{\JMU } 

\newcommand*{\KYUNGPOOK }{ Kyungpook National University, Daegu 702-701, South Korea} 
\affiliation{\KYUNGPOOK } 

\newcommand*{\MIT }{ Massachusetts Institute of Technology, Cambridge, Massachusetts  02139-4307} 
\affiliation{\MIT } 

\newcommand*{\UMASS }{ University of Massachusetts, Amherst, Massachusetts  01003} 
\affiliation{\UMASS } 

\newcommand*{\UNH }{ University of New Hampshire, Durham, New Hampshire 03824-3568} 
\affiliation{\UNH } 

\newcommand*{\NSU }{ Norfolk State University, Norfolk, Virginia 23504} 
\affiliation{\NSU } 

\newcommand*{\OHIOU }{ Ohio University, Athens, Ohio  45701} 
\affiliation{\OHIOU } 

\newcommand*{\ODU }{ Old Dominion University, Norfolk, Virginia 23529} 
\affiliation{\ODU } 

\newcommand*{\PENN }{Penn State University, University Park, Pennsylvania 16802, USA}
\affiliation{\PENN } 

\newcommand*{\PITT }{ University of Pittsburgh, Pittsburgh, Pennsylvania 15260} 
\affiliation{\PITT } 

\newcommand*{\ROMA }{ Universita' di ROMA III, 00146 Roma, Italy} 
\affiliation{\ROMA } 

\newcommand*{\RPI }{ Rensselaer Polytechnic Institute, Troy, New York 12180-3590} 
\affiliation{\RPI } 

\newcommand*{\RICE }{ Rice University, Houston, Texas 77005-1892} 
\affiliation{\RICE } 

\newcommand*{\URICH }{ University of Richmond, Richmond, Virginia 23173} 
\affiliation{\URICH } 

\newcommand*{\SCAROLINA }{ University of South Carolina, Columbia, South Carolina 29208} 
\affiliation{\SCAROLINA } 

\newcommand*{\JLAB }{ Thomas Jefferson National Accelerator Facility, Newport News, Virginia 23606} 
\affiliation{\JLAB } 

\newcommand*{\UNIONC }{ Union College, Schenectady, NY 12308} 
\affiliation{\UNIONC } 

\newcommand*{\VT }{ Virginia Polytechnic Institute and State University, Blacksburg, Virginia   24061-0435} 
\affiliation{\VT } 

\newcommand*{\VIRGINIA }{ University of Virginia, Charlottesville, Virginia 22901} 
\affiliation{\VIRGINIA } 

\newcommand*{\WM }{ College of William and Mary, Williamsburg, Virginia 23187-8795} 
\affiliation{\WM } 

\newcommand*{\YEREVAN }{ Yerevan Physics Institute, 375036 Yerevan, Armenia} 
\affiliation{\YEREVAN } 

\newcommand*{\deceased }{ Deceased} 
%\affiliation{\deceased } 

%\newcommand*{\NONE }{ unknown} 
%\affiliation{\NONE } 

\newcommand*{\NOWNCATU }{ North Carolina Agricultural and Technical State University, Greensboro, NC 27411}

\newcommand*{\NOWECOSSEG }{ University of Glasgow, Glasgow G12 8QQ, United Kingdom}

\newcommand*{\NOWSCAROLINA }{ University of South Carolina, Columbia, South Carolina 29208}

\newcommand*{\NOWJLAB }{ Thomas Jefferson National Accelerator Facility, Newport News, Virginia 23606}

\newcommand*{\NOWOHIOU }{ Ohio University, Athens, Ohio  45701}

\newcommand*{\NOWFIU }{ Florida International University, Miami, Florida 33199}

\newcommand*{\NOWODU }{ Old Dominion University, Norfolk, Virginia 23529}

\newcommand*{\NOWINFNFR }{ INFN, Laboratori Nazionali di Frascati, Frascati, Italy}

\newcommand*{\NOWCMU }{ Carnegie Mellon University, Pittsburgh, Pennsylvania 15213}

\newcommand*{\NOWCUA }{ Catholic University of America, Washington, D.C. 20064}

\newcommand*{\NOWINDSTRA }{ Systems Planning and Analysis, Alexandria, Virginia 22311}

\newcommand*{\NOWASU }{ Arizona State University, Tempe, Arizona 85287-1504}

\newcommand*{\NOWCISCO }{ Cisco, Washington, DC 20052}

\newcommand*{\NOWUK }{ University of Kentucky, LEXINGTON, KENTUCKY 40506}

\newcommand*{\NOWSACLAY }{ CEA-Saclay, Service de Physique Nucl\'eaire, F91191 Gif-sur-Yvette,Cedex, France}

\newcommand*{\NOWRPI }{ Rensselaer Polytechnic Institute, Troy, New York 12180-3590}

\newcommand*{\NOWDUKE }{ Duke University, Durham, North Carolina 27708-0305}

\newcommand*{\NOWUNCW }{ North Carolina}

\newcommand*{\NOWHAMPTON }{ Hampton University, Hampton, VA 23668}

\newcommand*{\NOW }{ }

\newcommand*{\NOWTulane }{ Tulane University, New Orleans, Lousiana  70118}

\newcommand*{\NOWGEORGETOWN }{ Georgetown University, Washington, DC 20057}

\newcommand*{\NOWJMU }{ James Madison University, Harrisonburg, Virginia 22807}

\newcommand*{\NOWURICH }{ University of Richmond, Richmond, Virginia 23173}

\newcommand*{\NOWCALTECH }{ California Institute of Technology, Pasadena, California 91125}

\newcommand*{\NOWMOSCOW }{ Moscow State University, General Nuclear Physics Institute, 119899 Moscow, Russia}

\newcommand*{\NOWVIRGINIA }{ University of Virginia, Charlottesville, Virginia 22901}

\newcommand*{\NOWYEREVAN }{ Yerevan Physics Institute, 375036 Yerevan, Armenia}

\newcommand*{\NOWRICE }{ Rice University, Houston, Texas 77005-1892}

\newcommand*{\NOWINFNGE }{ INFN, Sezione di Genova, 16146 Genova, Italy}

\newcommand*{\NOWROMA }{ Universita' di ROMA III, 00146 Roma, Italy}

\newcommand*{\NOWBATES }{ MIT-Bates Linear Accelerator Center, Middleton, MA 01949}

\newcommand*{\NOWKYUNGPOOK }{ Kungpook National University, Taegu 702-701, South Korea}

\newcommand*{\NOWFSU }{ Florida State University, Tallahassee, Florida 32306}

\newcommand*{\NOWVSU }{ Virginia State University, Petersburg,Virginia 23806}

\newcommand*{\NOWORST }{ Oregon State University, Corvallis, Oregon 97331-6507}

\newcommand*{\NOWGWU }{ The George Washington University, Washington, DC 20052}

\newcommand*{\NOWMIT }{ Massachusetts Institute of Technology, Cambridge, Massachusetts  02139-4307}

%%%%%%%%%%%%%%%%%%%% authors %%%%%%%%% 

\author{D.~Protopopescu}
  \email{protopop@jlab.org}
  \altaffiliation[Present affiliation:]{\ECOSSEG}
  \affiliation{\UNH} 
\author{F.~W.~Hersman}
   \affiliation{\UNH }
\author{M.~Holtrop}
   \affiliation{\UNH }
\author{G.~Adams}
     \affiliation{\RPI}
\author{P.~Ambrozewicz}
     \affiliation{\FIU}
\author{E.~Anciant}
     \affiliation{\SACLAY}
\author{M.~Anghinolfi}
     \affiliation{\INFNGE}
\author{B.~Asavapibhop}
     \affiliation{\UMASS}
\author{G.~Asryan}
     \affiliation{\YEREVAN}
\author{G.~Audit}
     \affiliation{\SACLAY}
% ...
\collaboration{The CLAS Collaboration}
     \noaffiliation

%%%%%%%%%%%% End new author list %%%%%%%%%%%%%%%%%%%%

%and theorists
%\author{D.~Debruyne}
%     \affiliation{\GENT}
%\author{J.~Ryckebusch}
%     \affiliation{\GENT}			

%%%%%%%%%%%%%%%%%%%%%%%%%%%%%%%%%%%%%%%%%

\date{\today}

\begin{abstract}

Single spin azimuthal asymmetries $A_{LT'}$ were measured at Jefferson
Lab using 2.2 and  4.4 GeV longitudinally polarised electrons incident
on $^{4}$He  and $^{12}$C targets  in the CLAS detector.  $A_{LT'}$ is
related   to   the  imaginary   part of    the longitudinal-transverse
interference   and in  quasifree   nucleon  knockout   it provides  an
unambiguous signature for final state interactions (FSI). Experimental
values of $A_{LT'}$ were found to be below $5\%$, typically $|A_{LT'}|
\leq 3\%$ for data with good statistical  precision.  Optical Model in
Eikonal  Approximation   (OMEA)  and Relativistic  Multiple-Scattering
Glauber Approximation (RMSGA) calculations are  shown to be consistent
with the measured asymmetries.

\end{abstract}

%Check these on: http://publish.aps.org/eprint/gateway/pacslist

\maketitle

\section{Introduction}

Although  quasielastic ($\omega \approx Q^2/2m_p$) $(e,e'p)$ reactions
have  been  a  well-used  tool  for the   study  of nuclear  structure
\cite{frullani,kelly1,lapikas93,steenhoven91,deWitt90}  and   received
considerable      attention          from            the     theorists
\cite{mahaux91,dickhoff92,boffi93}, gaps  in our understanding of this
reaction still persist.  One issue that stands out is the necessity of
a   thorough  understanding   of   final  state  interactions   (FSI).
Understanding FSI is a crucial ingredient in  being able to understand
short-range correlations in  nuclei \cite{Zhang:PhD}, effects  such as
the transition to  a quark-gluon picture  at higher $Q^2$, and nuclear
transparency  \cite{Lapikas:1999ss, Lava04, Fissum:2004we}. It is also
important  in   testing  the   predictions  of perturbative    QCD and
understanding dense nuclear matter.

There    are   numerous approaches   to   FSI:  distorted wave impulse
approximation (DWIA),   the  eikonal  approximation,  Glauber  theory,
etc. \cite{Debruyne:2001yi}, but  exact calculations are only possible
for $A \le 3$ at low  momentum.  The effects of  FSI must therefore be
measured experimentally, but they   cannot be isolated directly   in a
cross-section measurement.  They can  only be identified through their
interference  with the   dominant process,  i.e.   by separating   the
interference terms  of the  polarised cross-section \cite{RD:86}. This
is now possible following the  commissioning of reliable, high-current
polarised  sources at facilities such as  Mainz, MIT-Bates, NIKHEF and
TJNAF.

The beam  helicity  asymmetry $A_{LT'}$ turns    out to be   the ideal
observable for the study of FSI.  In general, $A_{LT'}$ corresponds to
the imaginary  part   of   the   longitudinal-transverse  interference
component of the  hadron tensor and  it vanishes whenever the reaction
proceeds     through  a channel     with  a   single  dominant   phase
\cite{boffi_c,co}.   Indeed, in the  plane wave  impulse approximation
(PWIA) where the rescattering is ignored and the hadron tensor is real
and symmetric, $A_{LT'}$ is zero.

In quasifree proton knockout, $A_{LT'}$ is almost  entirely due to the
interference of the   the two dominant  channels:  direct knockout and
rescattering through FSI.  Because $A_{LT'}$ is much less sensitive to
other effects such as meson-exchange  currents (MEC), it is considered
the best  observable for monitoring  rescattering effects  in knockout
reactions \cite{boffi_book}.

A measurement  of $A_{LT'}$ requires  a polarised  electron beam. When
polarised beam is used,  the  differential cross-section contains  two
terms  \cite{RD:86}:   a  helicity-independent term    $\Sigma$  and a
helicity-dependent term $h\Delta$,  with $h$ standing for the electron
helicity  $h=\pm1$.  They can  be separated in  the helicity asymmetry
$A_{LT'}$ defined as
\begin{equation}
\label{as1} A_{LT'}=\frac{d\sigma^+ - d\sigma^-}{d\sigma^+ + d\sigma^-}
= \frac{\Delta}{\Sigma}
\end{equation}
and measured  by simply flipping the beam  helicity.  In this formula,
the $d\sigma^+$  and  $d\sigma^-$ denote  differential  cross-sections
corresponding to  the +1 and  --1 helicities,  respectively. Practical
advantages    of  extracting    this observable    are  that  detector
efficiencies  cancel out in the ratio  and  that spectroscopic factors 
are not required for comparison with theory.

Two previous measurements of $A_{LT'}$ on $^{12}$C were carried out at
MIT-Bates  by Mandeville  et  al.  \cite{Mandeville:1994, dolfini} and
Jiang et  al. \cite{Jiang}.  However, while  the data were  seen to be
consistent with  DWIA,    they were  too  limited   to   draw  further
conclusions.  More  data,   with   higher statistical  accuracy,   are
necessary to differentiate among the models.  To date, no measurements
of $A_{LT'}$ in $^4$He have been done although theoretical predictions
have  been   made by Laget     \cite{laget_94}.  $^4$He is interesting
because,  despite  being  a four-body   system,  it is  a high density
nucleus.   A comprehensive  analysis   of FSI effects  in $^4$He  will
provide good insight into the significance of these effects in heavier
nuclei.

This article presents a survey of $A_{LT'}$ asymmetries  in $(\vec e ,
e'p)$  reactions on $^{12}$C  and  $^4$He in the quasielastic  regime,
exploring    kinematics not  previously   accessible.   The  questions
addressed  are: what  is  the strength of    the asymmetry signal  and
whether     present   models    can  describe   qualitatively   and/or
quantitatively the measurements.  The  theoretical models used  herein
were proven to successfully reproduce  the $L$, $T$ and $LT$  response
functions   \cite{Debruyne:2001yi,  Lava04, Fissum:2004we}, but  their
accuracy in describing the $LT'$ term has not yet been tested.

\section{Kinematic variables and response functions}\label{kine}

\begin{figure}
\centering
\epsfig{figure=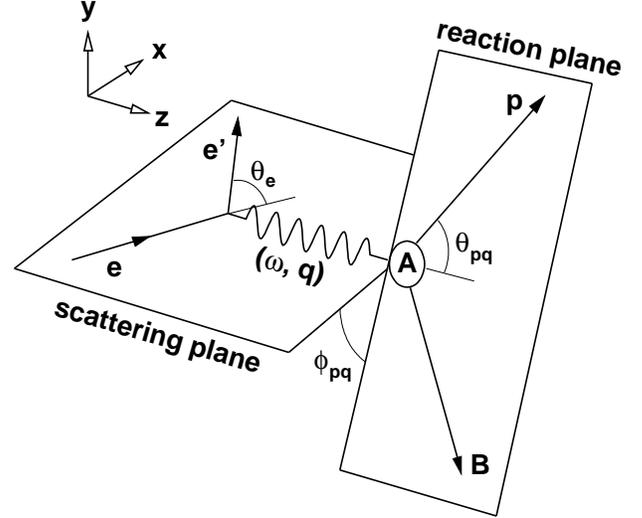,width=0.45\textwidth}
\caption{\label{e-scatt} Kinematics for the semi-exclusive 
A$(\vec e,e'p)$B reaction.}
\end{figure}

Electron scattering in the one-photon-exchange approximation (OPEA) is
schematically shown  in Fig.~\ref{e-scatt}.   An electron of initial
energy $E_e$  and  momentum  $\mathbf{k}$ scatters  through  an  angle
$\theta_e$ to a  final energy $E_e'$  and momentum $\mathbf{k'}$.  The
reaction plane  is rotated  by  an angle  $\phi_{pq}$ relative  to the
scattering plane. The  target nucleus is  denoted  by $A$ and  the
undetected recoiling  system by $B$.   The  ejected  proton  is   detected in
coincidence with  the scattered electron  $e'$.  The coordinate system
is  chosen such that the  $z$-axis  lies along  the momentum  transfer
$\mathbf{q}$ and the   $y$-axis is  perpendicular to  the   scattering
plane, parallel to  $\mathbf{k}\times\mathbf{k'}$.  For  $Q^2$ we  use
the   convention  $Q^2 =   -q_\nu  q^\nu \geq   0$,  where  $q$ is the
four-momentum of the virtual photon.

%%%%%%%%%%%%%%%%%%%%%%%%%%%%% e-p scattering %%%%%%%%%%%%%%%%%%%%%%%%%%%%%%%
% Figure made with: kine-diagram.C
%%%%%%%%%%%%%%%%%%%%%%%%%%%%%%%%%%%%%%%%%%%%%%%%%%%%%%%%%%%%%%%%%%%%%%%%%%%%

The missing energy $E_m$ and momentum $p_m$ for this semi-exclusive channel 
can be reconstructed as:
\begin{equation}
\label{Emiss} E_m = \omega-T_p-T_r, \qquad\quad \mathbf{p}_m 
                  = \mathbf{q} - \mathbf{p},
\end{equation}
where  $\mathbf{q} = \mathbf{k} -  \mathbf{k'}$  and  $\omega = E_e  -
E_e'$   are  the momentum   and   energy transferred by  the electron,
respectively, $T_p$  is the kinetic  energy of the outgoing proton and
$T_{r}$ is the kinetic energy of the recoiling system (see Fig. 
\ref{e-scatt}). 

In the absence of detected initial or final hadronic-state polarisation,
the cross-section for polarised electron scattering in the laboratory 
frame can be expressed as \cite{RD:86}:  
\begin{eqnarray}
\frac{d^5\sigma}{dE_e' d\Omega_e d\Omega_p} 
&=& K \sigma_M (v_Tf_T + v_Lf_L   \label{cs4} \\ 
&+& v_{TT}f_{TT}\cos2\phi_{pq} \nonumber\\ 
&+& v_{LT}f_{LT}\cos\phi_{pq} \nonumber\\ 
&+& h v_{LT'}f_{LT'}\sin\phi_{pq}) \nonumber\ ,
\end{eqnarray}
where  $\sigma_M$  is the Mott  cross section   and $K$  includes the
phase-space and recoil factors. In  the ultra-relativistic limit,  the
electron helicity states $h=+1$ and $h=-1$ correspond to spin parallel
and anti-parallel to $\mathbf{k}$, respectively.

The   labels $L$  and $T$   refer to  the  longitudinal and transverse
components of the  virtual  photon polarisation and  therefore  
correspond  to the electromagnetic  current components with respect to
the direction    of    $\mathbf{q}$.   Double   subscripts    indicate
interference terms.  The coefficients $v_i$ ($i = T,L,TT,LT \text{ and }
LT'$) are known  functions  of  the electron kinematics. 
The  response functions $f_i$ are proportional
to bilinear combinations of  the  nuclear current matrix elements  and
contain all of the nuclear structure information.

From equations (\ref{as1}) and (\ref{cs4}), the single spin 
asymmetry $A_{LT'}(\phi_{pq})$ can be written as:
\begin{eqnarray}
A_{LT'}(\phi_{pq})  &=&  
v_{LT'}f_{LT'}\sin\phi_{pq}/[ f_d +\label{as2}\\
  & &  v_{TT}f_{TT}\cos2\phi_{pq} +
   v_{LT}f_{LT}\cos\phi_{pq}] \ ,\nonumber
\end{eqnarray}
where $f_d = v_Tf_T + v_Lf_L$ denotes the $\phi$-independent term.

\section{Experiment}

The data were taken during the e2a period between April 15 and May 27,
1999 in Hall B at Jefferson  Laboratory using polarised electron beams
of energy $E=2.261$ GeV  and $E=4.461$ GeV at  beam currents of $2-10$
nA.

The  electron beam at CEBAF  was produced by  a  strained GaAs crystal
optically  pumped     by      circularly    polarised   laser    light
\cite{laser}. This setup permits the rapid flipping of the sign   of
the  electron helicity.  Helicity  pulses were  associated in pairs of
opposite helicity and the  leading   pulse helicity  was chosen by   a
pseudo-random  number      generator    in    order    to    eliminate
helicity-correlated  fluctuations.    The  electron  polarisation  was
determined  by  frequent   M\o ller  polarimeter   measurements to  be
$\langle  P_{B}    \rangle    =    0.63\pm   0.02\mbox{(stat.)}    \pm
0.03\mbox{(syst.)} $.  The polarisation  measurements and the helicity
decoding procedure are described in \cite{helsign}.

The $^{12}$C target was  a $1\times 1$  cm$^2$ plate of 1~mm thickness
and density $\rho_C = 1.786$ g/cm$^3$.  The helium target consisted of
liquid    $^{4}$He  ($\rho_{He} =   0.139$   g/cm$^3$)  contained in a
cylindrical aluminium  cell. Two  cells  were  used during  our   data
taking: 1) $4.99$ cm long and $0.97$ cm in diameter and 2) $3.72$ cm
long and $2.77$ cm in diameter.

Final  state  particles were detected  in  the  CEBAF Large Acceptance
Spectrometer (CLAS) which is described in detail in \cite{CLAS}.  CLAS
consists of a superconducting  toroidal magnet  and drift chambers  to
reconstruct   the momenta of  the tracks  of charged particles between
8$^0$ and  142$^0$ in  polar  angle and  with roughly  80\%  azimuthal
coverage.    It  has    gas Cherenkov   counters    (CC)  for electron
identification, scintillation    counters  for   measurement   of  the
time-of-flight (TOF) and electromagnetic calorimeters (EC) to identify
showering particles such as   electrons and photons.  The  trigger was
formed by fast coincidences between the CC and  EC at 2.261 GeV and by
EC  alone at  4.461 GeV.   Data were taken  at  an acquisition rate of
approximately 2.2  kHz and an  average (nucleon) luminosity during the
run of ${\cal L}=7\times 10^{33}$ s$^{-1}$cm$^{-2}$.

\section{Data Reduction and Analysis}\label{analysis}

Data    were  acquired  with  the   CLAS   ``single electron trigger''
configuration described above.   Further improvements in the  electron
identification  were   performed  offline,  where  the  total  energy
deposited  in  EC  was  used   to  identify  electrons and   eliminate
background such    as  negative pions.    Protons  were  identified by
time-of-flight.   The    reconstructed electron  and   proton   vertex
coordinates  were used to eliminate events  originating in the windows
and temperature  shield of the liquid-target cells. Electron momentum
corrections were applied to compensate for drift chamber misalignments
and uncertainties in   the  magnetic field  mapping. These  corrections
were in the range of 1--3\%.

With CLAS, it  is customary to  use software fiducial cuts to  exclude
regions of non-uniform detector response.  However, we will show later
on in this paper how these detector responses cancel  out in the ratio
Eq. (\ref{as1}). For that reason no fiducial cuts are deemed necessary
for  the  present analysis.   Also,  since the statistics  are low and
integration over a relatively large  missing energy range is employed,
we considered it  unnecessary  to apply  radiative corrections on  the
data. The systematic error introduced on $A_{LT'}$ by this omission is
estimated to be  below 2\%. This figure is  based on calculations done
for \cite{Joo03} using the program {\small EXCLURAD} \cite{EXCLURAD}.

Data  recorded during the run  \footnote{CLAS torus magnet current was
set at 2500  A}  amounted to  323 M triggers   at 2.261 GeV and  346 M
triggers at 4.461 GeV for carbon, and 310  M triggers at 2.261 GeV and
442 M triggers at 4.461 GeV for  helium.  After data reduction and the
quasielastic cut described  below, the statistics corresponding to the
two beam settings  were 3.8 M and 0.3  M for carbon, respectively, and
2.7 M and 0.25 M for helium, respectively.

The quasielastic (QE) kinematics region was selected by a cut along the
quasielastic ridge  $\omega = Q^2/2m_p +  \Delta E$ with the condition
$\omega_1 < \omega < \omega_2$ and limits given by the equation:
\begin{equation}
\label{QEcut} \omega_{1,2} = (1/2 \pm \xi)Q^2/m_p + \Delta E ,
\end{equation}
where $\xi=0.2$ sets the width  of our cut and  $\Delta E$ is a  shift
due to the momentum dependence of the nucleus-nucleon potential, taken
to be equal  to 0.03 GeV.  This cut  is roughly equivalent   to $0.7 < x_B <
1.6$,  where $x_B$  is  the  Bjorken   variable. In conjunction   with
(\ref{QEcut})  we  rejected events with  $E_m  >  0.1$~GeV   to  reduce  the
contribution of multi-particle final states.

The $Q^2$ coverage of the selected data was 0.35 -- 1.80 GeV$^2$/c$^2$
at $2.261$ GeV beam energy and 0.80 -- 2.40 GeV$^2$/c$^2$ at $4.461$ GeV. 
For the study of the   quasielastic regime we divided the   kinematics
explored into four $Q^2$ bins, nine $\,p_m$ bins and sixty $\phi_{pq}$
bins. The cut  from (\ref{QEcut}) is applied  for each case so that in 
practise we work with 4-dimensional  bins in coordinates $Q^2$, $p_m$,
$\omega$ and $\phi_{pq}$.

For  the investigation of the missing  energy  dependence of $A_{LT'}$
outside the quasielastic regime, we used 0.1 GeV bins  in $E_m$ and we
integrated over   $Q^2$  while splitting  $p_m$  into two  large bins.
Consequently, we worked with  3-dimensional bins in coordinates $E_m$,
$p_m$ and $\phi_{pq}$.

To measure the beam helicity asymmetry defined in Eq. (\ref{as1})
we calculated the ratio
\begin{equation}
\label{as3} A_m = \frac{N^+ - N^-}{N^+ +  N^-} ,
\end{equation}
where $N^+$  and $N^-$   are,  respectively, the number of   $(e,e'p)$
events with positive and negative helicity within the chosen kinematic
bin.  

False asymmetries were investigated by assigning helicities randomly to
each event and comparing the resulting $A_m^{random}$ asymmetry with zero.
The false asymmetry values were much smaller than the statistical errors.

The yields $N^+$  and $N^-$ from (\ref{as3})  were corrected  for beam
charge asymmetry (BCA).  Since the  BCA corrections are extracted from
inclusive   data,  the  systematic   uncertainty  introduced  by  this
procedure is  approximately 1\%  of  the statistical error on  $N^\pm$
\cite{mythesis}.

\begin{figure}[t!]
\centering            
\epsfig{figure=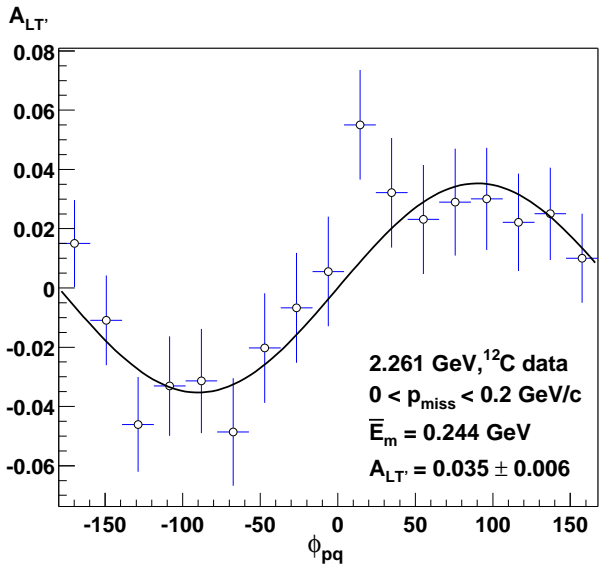, width=0.5\textwidth}
\caption{\label{sinusfit}  (Color online) Fit using function Eq. 
(\ref{fitf1}):  $^{12}$C data at  2.261   GeV are integrated over  all
$Q^2$ and over missing momentum in the interval $0  < p_m < 0.2$ GeV/c
(see  Fig.\ref{emdep_fig}.e).  For clarity,   the  number of  bins  is
reduced here  and the  asymmetry  is corrected  for beam  polarisation
using Eq. (\ref{as4}).}
\end{figure}

This procedure described in   the previous paragraph was  repeated for
all $60$ bins in $\phi_{pq}$.  Then these  values were used to extract
$A_{exp}$ by fitting the $\phi_{pq}$ dependence with the one-parameter
function:
\begin{equation}
\label{fitf1} A_m(\phi_{pq}) = A_{exp} \sin\phi_{pq} \ ,
\end{equation}
The $TT$ and $LT$ contributions  were  neglected in  this fit but  the
systematic uncertainty introduced  by this approximation was estimated
to be below 5\% \cite{mythesis}. A  fit with function (\ref{fitf1}) is
shown in Fig. \ref{sinusfit}.

%%%%%%%%%%%%%%%%%%%%%%%%%%%%%%%%%%%%%%%%%%%%%%%%%%%%%%%%%%%%%%%%%%%%%%
% Figure made with hist2eps.C
%%%%%%%%%%%%%%%%%%%%%%%%%%%%%%%%%%%%%%%%%%%%%%%%%%%%%%%%%%%%%%%%%%%%%%

For  small    measured asymmetries  the   statistical  uncertainty  on
$A_m(\phi_{pq})$   from   (\ref{as3}) was   dominated   by  the  $1/N$
dependence (where $N = N^+ + N^-$):
\begin{equation}
\label{as4e} \delta A_m(\phi_{pq}) \approx 
\sqrt{(1 - A_m^2(\phi_{pq}))/N} \approx 1/\sqrt{N} \ ,
\end{equation}
The MINUIT  \cite{min95} fit  uncertainty for $A_{exp}$  dominated the
statistical uncertainty on $A_m(\phi_{pq})$, which was in the range of
$\pm$20\%.

To  account for  the partial  polarisation  of the incident beam,  the
extracted asymmetry $A_{exp}$   was  scaled by the beam   polarisation
$P_B$ to obtain the true asymmetry $A_{LT'}$:
\begin{equation}
\label{as4} A_{LT'} = A_{exp}/P_B \ ,
\end{equation}
The beam polarisation was measured with a relative uncertainty of less
than 5\%.    This,  added in  quadrature  to  the   other sources   of
systematic errors mentioned before (radiative effects,  BCA and choice
of fit  function),  yield a systematic error  $<8$\%.   Thus, the data
presented herein is dominated by the fit uncertainty which is the only
one shown on the plots.

\section{Theoretical Calculations}

The  experimental    results are  compared   with  theoretical  models
developed  by  the  Ghent group  \cite{Dimi:2001,Debruyne:2001yi,dimi}.
Both models employ bound-state  wavefunctions  calculated within
the context of a mean-field approximation to the $\sigma-\omega$ model
\cite{so1,so2}.   In all   calculations, the  $\Gamma_{cc2}$  hadronic
current operator and the Coulomb gauge were used and the spectroscopic
factors for $^{12}$C were   set equal to  $S  = 2j +1$.   The electron
distortion - i.e. the distortion of the electron wave function in the
Coulomb field of the nucleus - has
been  neglected, since its effect for   nuclei as light  as $^4$He and
$^{12}$C is   very  small.    All  calculations presented    here  are
relativistic   and   include  the  spinor distortion  \footnote{Spinor
distortion  :  The lower (i.e.  small)  components  in the  Dirac wave
functions for  the nucleons take  on the following form $(\sigma \cdot
\mathbf{p})/(E + M + S(r) - V(r)) \times (radial\ wave\ function)$. By
"spinor distortion" one refers  to the presence of  the term "$S(r)  -
V(r)$", where $S(r)$ is the  scalar and $V(r)$  the vector part of the
"mean-field"  potential.} of  the lower components  in the bound-state
wave  function   due  to the    scalar and  vector   potential of  the
$\sigma-\omega$ model.

The Optical Model in the Eikonal Approximation  (OMEA) is based on the
relativistic  eikonal  approximation for  the ejectile scattering wave
function and a  relativistic  mean-field approximation to  the Walecka
model \cite{Debruyne:2000wh}.  The  optical potential used in the OMEA
calculations is  EDAIC, of  Cooper  $et\ al.$ \cite{cooper}. The  only
basic  difference   between OMEA  and   the traditional distorted-wave
impulse  approximation (DWIA)   \cite{boffi_book}  is the  use  of the
eikonal approximation to   construct the scattering wave   function in
OMEA.  While in DWIA the exact  solutions to the equations determining
the scattering wave function are used, the OMEA adopts the approximate
eikonal solutions to  the  same  equation  \cite{Debruyne:2001yi}. The
eikonal solution  approaches the exact one  if the missing momentum of
the ejectile   is sufficiently small  in  comparison with the momentum
transfer $q$.

An extension of  the eikonal method  is  introduced to cope with  high
proton   kinetic  energies ($T_p  \geq  1\,$GeV)  where the elementary
proton-nucleon scattering  becomes   highly inelastic and  the  use of
optical  potentials  is no  longer a natural  choice  for describing
final-state  interactions.       The  Relativistic Multiple-Scattering
Glauber Approximation (RMSGA) is  a relativistic generalisation of the
Glauber approach  described  in \cite{Ryckebusch:2003fc}.  Whereas the
optical potential approaches model the  FSI effects through estimating
the  loss of  flux from  elastic   proton-{\bf nucleus} data,  Glauber
approaches rely on elastic  proton-{\bf  nucleon} data.   The  Glauber
model   is  an  A-body  multiple-scattering theory,    which relies in
addition to  the eikonal on the so-called  frozen approximation.  This
requires the spectator nucleons to  be stationary during the time that
the ejectile travels through the target nucleus.

We investigate here missing momenta up to 0.6 GeV/c. However, we would
like to note that, as illustrated by the measurements of Liyanage $et\
al.$   \cite{nilanga}, at missing momenta above   0.3 GeV/c it is very
likely  that more complicated   reaction mechanisms (channel coupling,
two-nucleon knockout, etc.) come into play and can contribute by up to
50\%  of the   cross-section \cite{Fissum:2004we}. Comparisons between
theory  and data  above  $p_m = 0.3$ GeV/c   should be viewed in  this
context.

\section{Acceptance and Bin Averaging Effects}\label{etcomp}

Bin averaging effects are not  negligible when summing over relatively
wide  kinematics.  This was   a concern in  the  case of our $Q^2$ and
$\omega$ coordinates (see  section \ref{analysis}) since  the bins are
$0.36$ to $0.40$  GeV$^2$/c$^2$ wide  in  $Q^2$ and  about $50$  MeV  in
$\omega$.  These widths were  chosen in order   to keep the  statistical
errors reasonably small, but then the  theory calculations must account
for the resulting bin averaging, as described in the following.

Let us consider an arbitrary bin $B_n$
of width  $\Delta\chi = \Delta Q^2\Delta\omega \Delta p_m \Delta\phi_{pq}$ 
and denote by $\langle\chi\rangle_n$ its centre of weight. 
Let us observe that the r.h.s. of Eq. (\ref{as3}) can be put
in the form below:
\begin{equation}
\label{binning1}
A_m(\langle\chi\rangle_n) = 
%\frac{N^+ - N^-}{N^+ +  N^-} = 
\frac{\sum\limits_{k=1}^K (N^+_{k} - N^-_{k})}
{\sum\limits_{k'=1}^K (N^+_{k'} + N^-_{k'})} \equiv
\frac{\sum\limits_{k=1}^K \sigma^h_{k} \epsilon_k 
}{\sum\limits_{k'=1}^K \sigma^0_{k'} \epsilon_{k'}},
\end{equation}
where we formally divided the   $\,\omega\times Q^2$ subspace within bin  
$B_n$ into $K$   sub-bins and  denoted  by  $\epsilon_k$  the  (unknown) detector
acceptance   within   sub-bin $b_k$.  We   used  the  shorthand notations
$\sigma^h = \sigma^+ - \sigma^- $ and $\sigma^0 = \sigma^+ + \sigma^-$
for  the polarised and unpolarised  parts  of the total cross-section,
respectively.

After some manipulation, Eq. (\ref{binning1}) can be written as
\begin{eqnarray}
A_m(\langle\chi\rangle_n) &=& \sum\limits_{k=1}^K \left(
\frac{\sigma^h_{k}}{\sigma^0_{k}} \right)\left(
\frac{\sigma^0_{k} \epsilon_k }{
\sum\limits_{k'=1}^K \sigma^0_{k'} 
\epsilon_{k'}} \right)   \nonumber\\
 &\equiv & \sum\limits_{k=1}^K A_{m}(\langle\chi\rangle_k)\, w_k,
\label{binning2}
\end{eqnarray}
where we separated the weights $w_k$ defined as
\begin{equation}
w_k = \frac{\sigma^0_{k} \epsilon_k }{
\sum\limits_{k'=1}^K \sigma^0_{k'}
\epsilon_{k'}} = \frac{N_k}{N}, \qquad N = N^+ + N^- ,
\end{equation}
which can be easily calculated using the experimental yields. Then
theory calculations can be done for each $k$ sub-bin and added up 
within $B_n$, i.e.
\begin{equation}
A_t(\langle\chi\rangle_n) = 
\sum\limits_{k=1}^K A_{c}(\langle\chi\rangle_k) w_k ,
\label{binning3}
\end{equation}
where  $A_{c}(\langle\chi\rangle_k)$  are pure  model calculations for
the kinematics  $\langle\chi\rangle_k$.

To   determine the optimal number   of sub-bins,  $K$,  a compromise was
sought  between  a small number that   minimises  both the statistical
error  on $w_k$ and the  computational time, and   a large number that
better compensates for kinematic averaging. The value fulfilling these
conditions was found to be $K=4$  with negligible statistical error on
$w_k$ \cite{mythesis} .

Smooth  theoretical  curves are    obtained   by cubic  spline
interpolation of the $A_{t}(\langle\chi\rangle_n)$ array.

\section{Results}

Two variations of the $A_{LT'}$ asymmetry are presented: {\bf A)}
the missing momentum dependence at quasielastic kinematics with $E_m <
0.1$ GeV and {\bf B)} the missing energy dependence  up to $E_m = 0.9$
GeV.  Theoretical calculations  accompany the  missing  momentum
plots.

The $^{4}$He data is consistent with  zero except for a possible small
positive region at missing momentum around $0.15 < p_m  < 0.3$ at $Q^2
< 1.4$ GeV. The $^{12}$C 2.262 GeV data are  negative for $0.1 < p_m <
0.2$ and become mostly consistent  with zero for missing momenta above
$0.3$ GeV and for all $p_m$ range at 4.462 GeV.

\subsection{Missing momentum dependence}\label{pmiss_s}

\begin{figure}[t!]
\centering            
\epsfig{figure=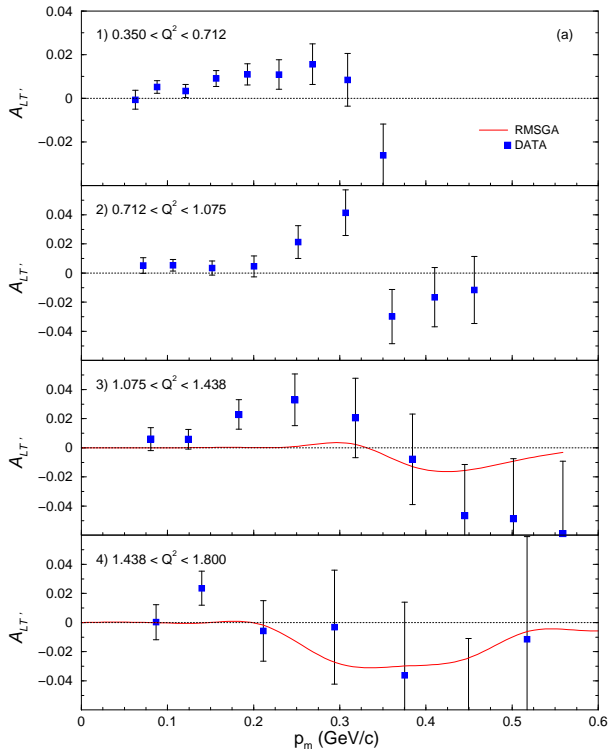,
	width=0.44\textwidth}
\epsfig{figure=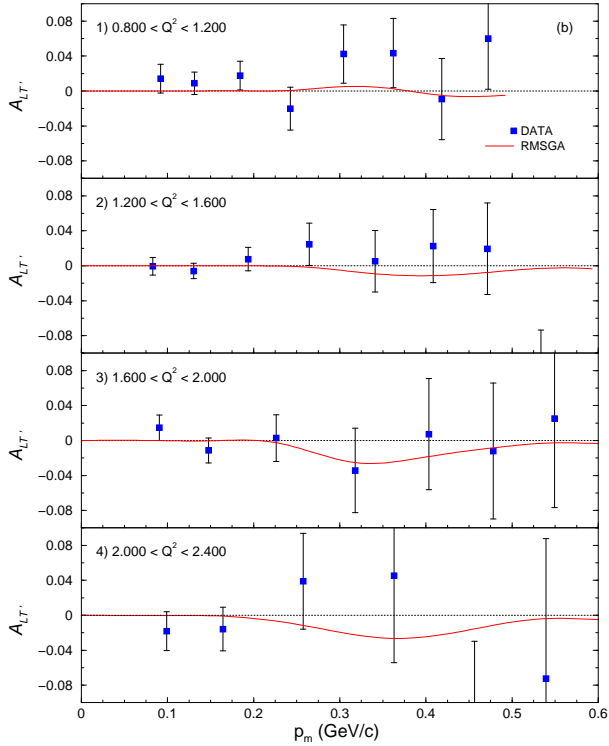,
	width=0.44\textwidth}
\caption{\label{he4_pmdep}  (Color online) Dependence of $A_{LT'}$            
versus missing momentum $p_m$ for  $^{4}$He$(e,e'p)$ at (a)  $E=$2.262
GeV and (b)  $E=$4.462 GeV, in four  $Q^2$ bins along the quasielastic
ridge.  The curves correspond  to RMSGA calculations used as described
in section \ref{etcomp}.  Error bars are statistical. }
\end{figure}

\begin{figure}[t!]
\centering            
\epsfig{figure=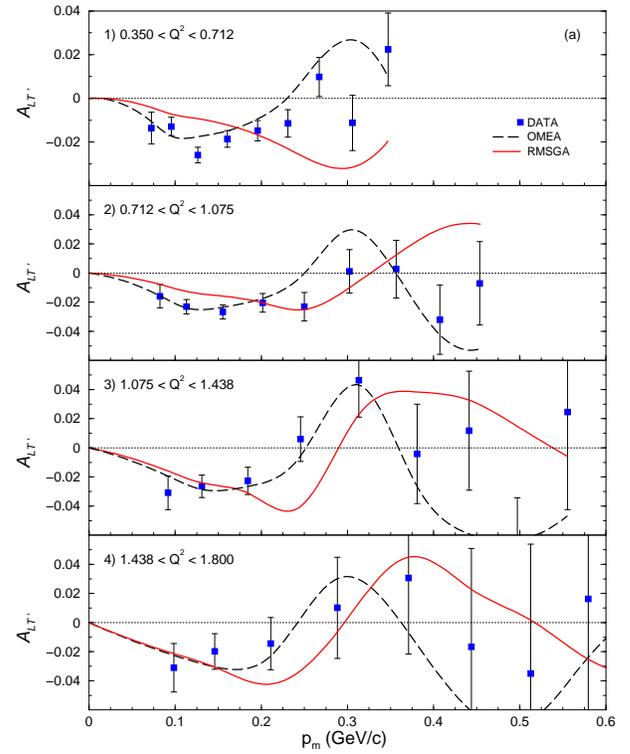,
	width=0.44\textwidth}
\epsfig{figure=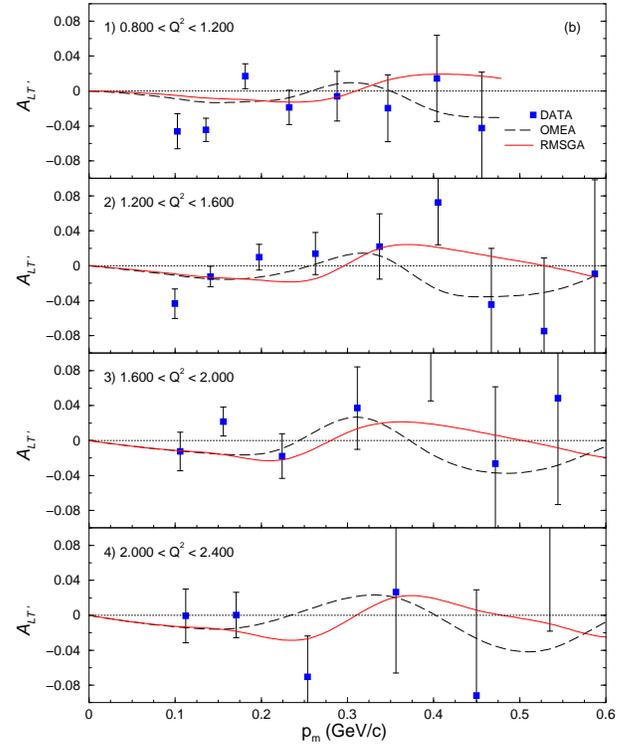,
	width=0.44\textwidth}
\caption{\label{c12_pmdep}  (Color online) Dependence of $A_{LT'}$            
versus missing momentum  $p_m$ for $^{12}$C$(e,e'p)$ at  (a) $E=$2.262
GeV and (b)  $E=$4.462 GeV, in  four $Q^2$ bins along the quasielastic
ridge.   The curves  correspond   to RMSGA (solid)  and OMEA  (dotted)
calculations used as  described in  section \ref{etcomp}.  Error  bars
are statistical. }
\end{figure}

The dependence of $A_{LT'}$ at  quasielastic kinematics as a  function
of $p_m$ is presented  in Fig.  \ref{he4_pmdep}--\ref{c12_pmdep}.  The
data  were divided into four $Q^2$  bins for each beam energy setting.
There  is a direct mapping  between $\theta_{pq}$ and $p_m$, such that
low  $p_m$ corresponds to  low  $\theta_{pq}$.  The equivalent angular
range  covered  by the mentioned plots  would  be $0^o < \theta_{pq} <
27^o$.

%%%%%%%%%%%%%%%%%%%%%%%%%%%%%%%%%%%%%%%%%%%%%%%%%%%%%%%%%%%%%%%%%%%%%%%%%%%
%%%%%%%%%%%%%%%%%%%%%%%%%%%%%%%%%%%%%%%%%%%%%%%%%%%%%%%%%%%%%%%%%%%%%%%%%%%
%%%%%%%%%%%%%%%%%%%%%%%%%%%%%%%%%%%%%%%%%%%%%%%%%%%%%%%%%%%%%%%%%%%%%%%%%%%
%%%%%%%%%%%%%%%%%%%%%%%%%%%%%%%%%%%%%%%%%%%%%%%%%%%%%%%%%%%%%%%%%%%%%%%%%%%

The  $^{12}$C  data  are accompanied   by  OMEA  calculations   at all
kinematics  and by RMSGA calculations  in the range $|\mathbf{q}| > 1$
GeV/c, where   the latter  model  is applicable.  The $^4$He   data is
compared with RMSGA  only.  The error bars  shown on the data  are fit
uncertainties as explained in section \ref{analysis}.

A first observation is that measured asymmetries are very small at low
$p_m$.   This is  to  be  expected as  low  $p_m$  corresponds  to low
$\theta_{pq}$  where  the   helicity   asymmetry  vanishes  with   the
phase-space:  $A_{LT'}   \sim \sin\theta_{pq}$.    The sharp   rise of
$A_{LT'}$ at low missing momentum shown by  the low $Q^2$ calculations
from  \cite{dolfini} is at our energy setting seen neither in
the measurement nor in the  calculations shown  herein. There
are at least  two  reasons for the  smaller  asymmetries  here: a)  at
higher energy the effects of FSI are smaller and the calculation would
naturally produce a  smaller  oscillation in  $A_{LT'}$ and  b) in the
case of  $^{12}$C   cancellation between   unresolved $1s$   and  $1p$
contributions significantly reduces the inclusive asymmetry.

The overall aspect of the plots is almost independent of $Q^2$. 
This is an important point to note when integrating over $Q^2$ in the 
next subsection.

Although  the statistical significance   is  not remarkably high,  the
2.262   GeV  data suggest  a   structure  between $0.1  < p_m   < 0.3$
GeV/c. Its  position appears to  not vary much with  $Q^2$ and, in the
case of $^{12}$C, both models more or less reproduce this feature.

The large change in slope seen on the $^{12}$C calculations at $p_m
\approx 0.3$ GeV/c   is likely due to  the  dip in the  $1s_{1/2}$ and
$1p_{3/2}$  momentum    distributions at  slightly   higher   $p_m$ as
calculated  in \cite{Ryckebusch:2003fc}.   This effect, attributed  to
bound-nucleon and ejectile  spinor distortion \cite{Fissum:2004we}, is
shifted to lower missing momenta and partially  washed out by the FSI.
The  difference between the two calculations  is  due to the fact that
RMSGA  will locate  the  effect at  higher  $p_m$  compared with OMEA,
reflecting  the fact that    the   predicted effect   of FSI    on the
observables  is  smaller   in  Glauber   approaches than   in  optical
potential-based models.

Although   carbon  data  (Fig.~\ref{c12_pmdep})  benefit  from  higher
statistical precision than the helium  data, separation of the $s$ and
$p$ shells in $^{12}$C is not possible due to the limited instrumental
resolution of CLAS (22 MeV FWHM).  The interplay between the two shells
is  important,  since our   calculations   showed that  the $s$-   and
$p$-shell     contributions        are    of    opposite          sign
\cite{mythesis,kelly-priv}.   The   effect  of  the $s$-shell  can  be
visualised by  comparing the 2.262 GeV  helium and  carbon plots: both
targets  feature the  mentioned structure  between $0.1  < p_m <  0.3$
GeV/c.  But while the $^4$He  asymmetry stays positive in this missing
momentum range, the $^{12}$C  asymmetry goes through  negative values,
most likely due to the interplay between the two shell components.

\subsection{Missing energy dependence}\label{emsect}

\begin{figure}
\centering     
\epsfig{figure=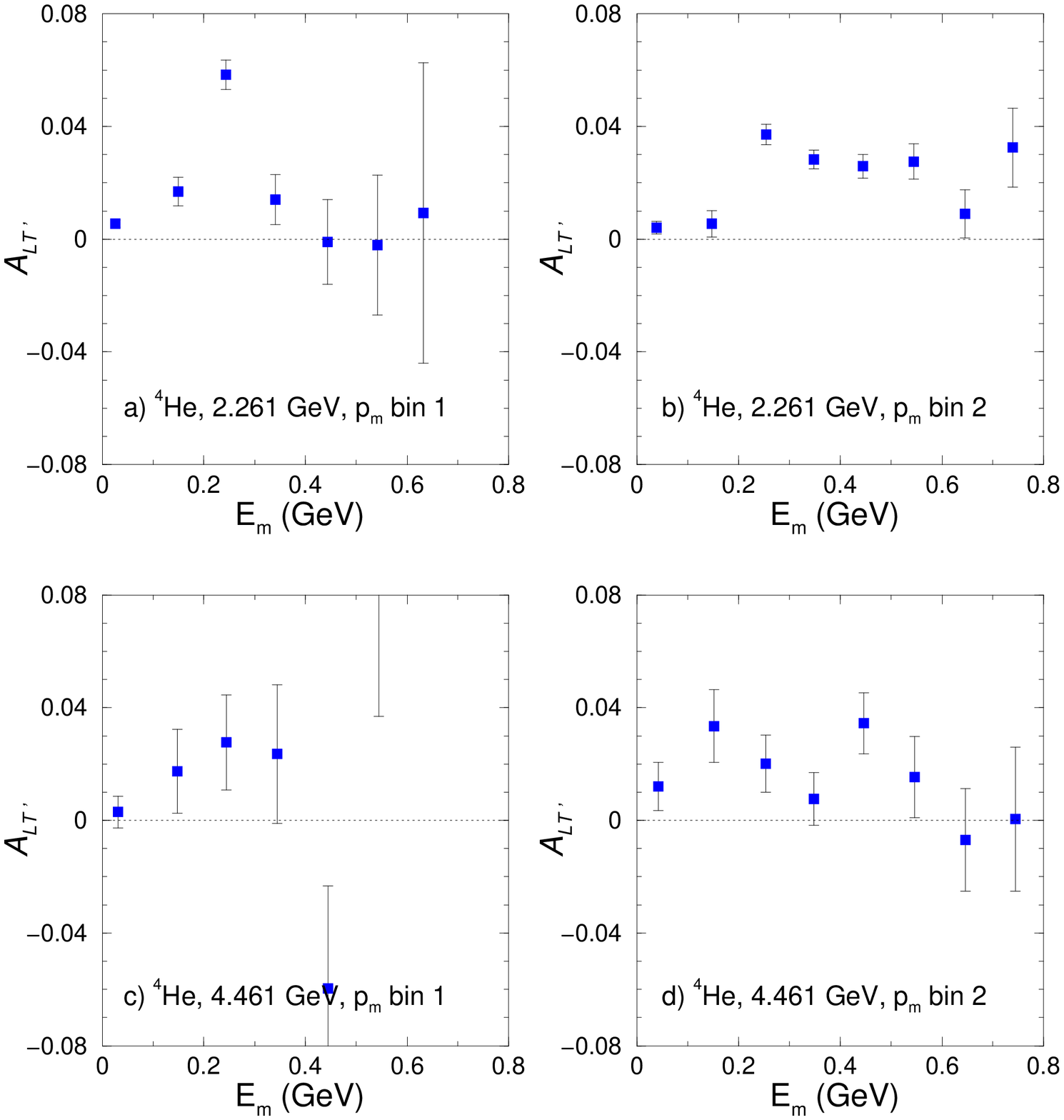,
	width=0.45\textwidth}
\epsfig{figure=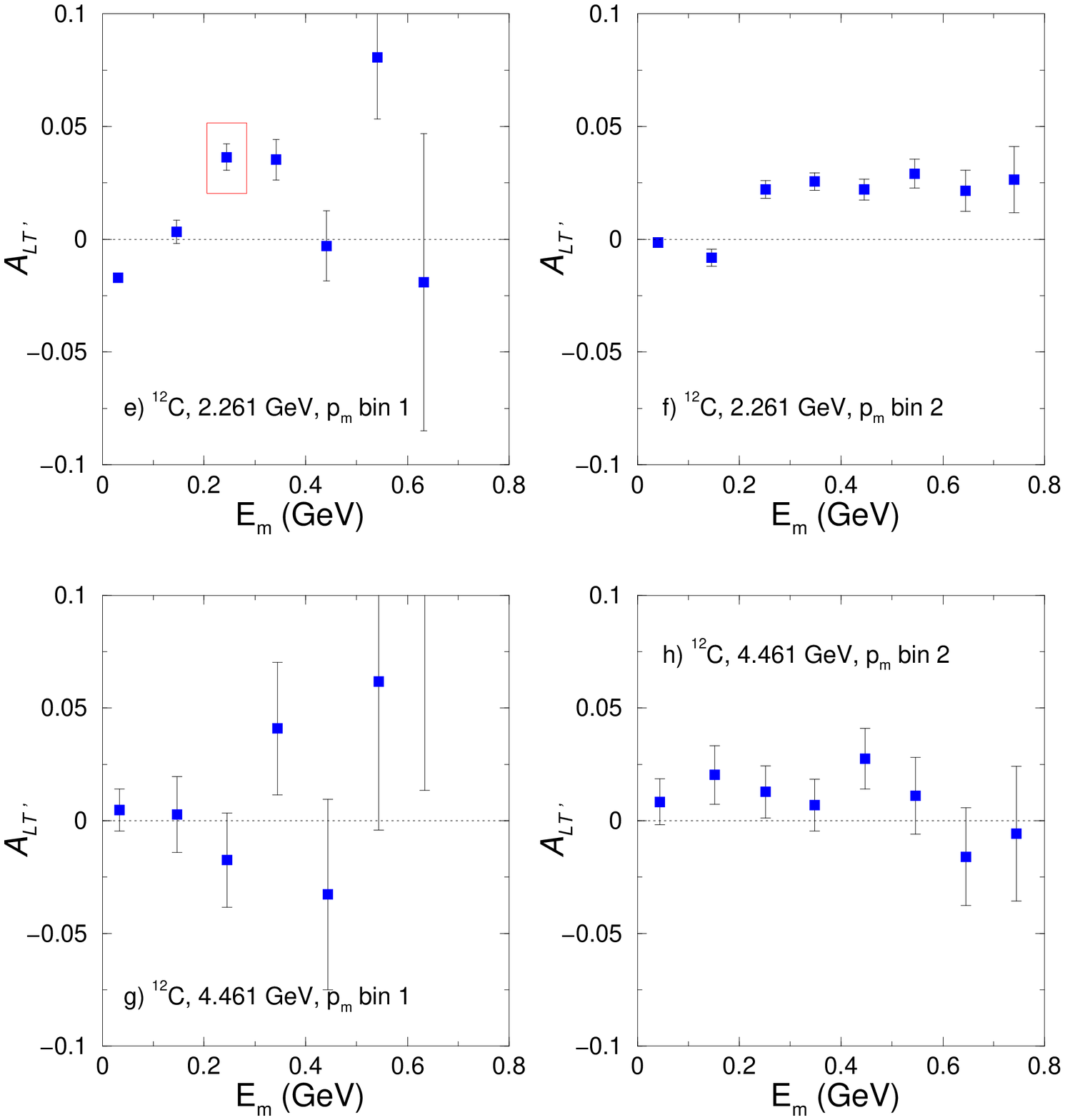,
	width=0.45\textwidth}
\caption{\label{emdep_fig} (Color online)
$A_{LT'}$  versus $E_m$  for helium  (plots   a--d) and carbon  (plots
e--h). The two missing  momentum bins are defined  as follows: 1)~$0 <
p_m < 0.2$ and 2)~$0.2  < p_m < 0.45$ with  $p_m$ in GeV/c. The  $E_m$
bins  are 0.1 GeV wide.  No  radiative corrections were applied. Error
bars    are  statistical.   The  boxed   point on plot   e  was
extracted from the sinusoid presented in Fig.\ref{sinusfit}.}
\end{figure}

The missing energy dependence of $A_{LT'}$  was further studied over a
range  spanning  up to  $E_m \leq 0.8$ GeV.   The  plots are shown  in
figures \ref{emdep_fig} (a--h). The very  first point on all the plots
corresponds to the  valence  knockout kinematics, investigated in  the
previous subsection, but integrated over a  larger $p_m$ interval (see
figure caption).

%%%%%%%%%%%%%%%%%%%%%%%%%%%%%% figure %%%%%%%%%%%%%%%%%%%%%%%%%%%%%%%%
%%%%%%%%%%%%%%%%%%%%%%%%%%%%%%%%%%%%%%%%%%%%%%%%%%%%%%%%%%%%%%%%%%%%%%

In accordance with the $s$-shell effect mentioned in
subsection \ref{pmiss_s}, the lowest $E_m$ sample point shows negative
asymmetries  for $^{12}$C compared   with positive asymmetries  in the
case of $^4$He (Fig. \ref{emdep_fig}, plots a and e).

There is a clear and dramatic increase in $A_{LT'}$ at $E_m = 0.25$ GeV
where the pion production channel opens.   It is very likely that this
extra channel interferes  with the  other  open channels to produce  a
larger  $A_{LT'}$. This  feature is  less  pronounced at higher  $p_m$
since here other processes besides direct knockout and pion production
are smearing the interference peak.

Theoretical calculations at higher missing energy to accompany these 
data would be valuable.

\section{Summary}

We have measured the $A_{LT'}$  asymmetry in quasielastic $(\omega 
\approx Q^2/2m_p)$ reactions on $^{12}$C and  $^4$He in 
order to determine the contribution of final state interactions in the
one-proton knockout reaction.

Overall,   the measured asymmetries in    the quasielastic region were
below  5\%, typically   in  the range  of  0-3\%  for  data with  good
statistics.  This seems to indicate that
FSI are  very small and unimportant   for these light nuclei  at these
energy  settings.  Nevertheless, the  results presented herein confirm
the presence    of  asymmetry  signals in $(\vec{e},e'p)$     data and
complement the  (scarce) world data  on this subject. Our survey could
indicate the  regions of interest  for  future investigations, such as
$p_m  \ge 0.3$~GeV/c  or higher  missing  energies, for which in  this
paper the calculations  either  differ or  were unavailable,  and  the
statistics used were relatively scarce.

While  areas   of  disagreement  were  observed,   all  models  showed
qualitative agreement with the experiment.  A detailed comparison with
theory  for $^{12}$C would require  separation  of major shells, since
they  are   predicted  to  contribute  with   opposite  sign  to   the
asymmetry. Within the range of validity of the models and summing over
the  final states, one can  assert that the theoretically modelled FSI
are consistent with the experiment.

\acknowledgments

The authors would like to thank R.~Owens for interesting discussions
and useful suggestions.  We would like  to acknowledge the outstanding
efforts of  the staff of the  Accelerator and the Physics Divisions at
JLab that made this experiment possible.

This work was  supported in part by  the Instituto Nazionale di Fisica
Nucleare, the French Centre National de la Recherche Scientifique, the
French Commissariat \`{a} l'Energie Atomique,  the U.S.  Department of
Energy, the National Science Foundation,  Emmy Noether grant from  the
Deutsche  Forschungsgemeinschaft,  the  Korean Science and Engineering
Foundation  and the  UK's  Engineering and  Physical Sciences Research
Council.   The Southeastern  Universities Research Association  (SURA)
operates the  Thomas  Jefferson National Accelerator  Facility for the
United States Department of Energy under contract DE-AC05-84ER40150.

\bibliography{refs}

\end{document}